\documentclass[twocolumn]{webofc}

\usepackage[varg]{txfonts}   
\usepackage{color}

\begin{document}
\title{Midrapidity cluster formation in heavy-ion collisions}
%
%

\author{\firstname{Elena} \lastname{Bratkovskaya}\inst{1,2,3}\fnsep\thanks{invited speaker,\email{E.Bratkovskaya@gsi.de}} 
    \and
        \firstname{Susanne} \lastname{Gl{\"a}ssel}\inst{4}\fnsep 
             \and
        \firstname{Viktar} \lastname{Kireyeu}\inst{3,5}\fnsep
           \and
        \firstname{J\"org} \lastname{Aichelin}\inst{6,7}\fnsep
           \and
        \firstname{Marcus} \lastname{Bleicher}\inst{2,3,1}\fnsep
           \and           
        \firstname{Christoph} \lastname{Blume}\inst{4}\fnsep
           \and
        \firstname{Gabriele} \lastname{Coci}\inst{2,3}\fnsep
           \and
        \firstname{Vadim} \lastname{Kolesnikov}\inst{5}\fnsep
           \and
        \firstname{Jan} \lastname{Steinheimer}\inst{7}\fnsep
           \and
        \firstname{Vadim} \lastname{Voronyuk}\inst{3,5}\fnsep
}

\institute{
GSI Helmholtzzentrum f\"ur Schwerionenforschung GmbH,
  Planckstr. 1, 64291 Darmstadt, Germany
\and
  Institut f\"ur Theoretische Physik, Johann Wolfgang Goethe-Universit\"at,
  Max-von-Laue-Str. 1, 60438 Frankfurt, Germany         
\and
Helmholtz Research Academy Hessen for FAIR (HFHF), GSI Helmholtz Center for Heavy Ion Physics. Campus Frankfurt, 60438 Frankfurt, Germany
 \and
Institut f\"ur Kernphysik, Johann Wolfgang Goethe-Universit\"at,
Max-von-Laue-Str. 1, 60438 Frankfurt, Germany
 \and 
 Joint Institute for Nuclear Research, Joliot-Curie 6, 141980 Dubna, Moscow region, Russia
  \and
SUBATECH, Universit\'e de Nantes, IMT Atlantique, IN2P3/CNRS
4 rue Alfred Kastler, 44307 Nantes cedex 3, France
 \and
Frankfurt Institute for Advanced Studies, Ruth Moufang Str. 1, 60438 Frankfurt, Germany
          }

\abstract{%
We study the production of clusters and hypernuclei at midrapidity employing
the Parton-Hadron-Quantum-Molecular-Dynamics (PHQMD) approach, a microscopic n-body transport model based on the QMD propagation of the baryonic degrees of freedom with density dependent 2-body potential interactions.
 In PHQMD the cluster formation occurs dynamically, caused by the interactions. The clusters are recognized by the Minimum Spanning Tree (MST) algorithm. 
We present the PHQMD results for cluster and hypernuclei formation  in comparison with the available experimental data at relativistic energies.
PHQMD allows to study the time evolution of formed clusters and the origin of their production, which helps to understand how such weakly bound objects are formed and survive in the rather dense and hot environment created in heavy-ion collisions. It offers therefore an explanation of the 'ice in the fire' puzzle.
To investigate whether this explanation of the 'ice in the fire' puzzle applies only to the MST results we study also the deuterons production by coalescence. We embed MST and coalescence  in the PHQMD and UrQMD transport approaches in order to obtain  model independent results.  We find that both clustering procedures give very similar results for the deuteron observables in the UrQMD as well as in the PHQMD environment. This confirms that our solution for the 'ice in the fire' puzzle is common to MST and coalescence  and independent of the transport approach.  
}
\maketitle
\section{Introduction}
\label{intro}

Cluster and hypernuclei production in heavy-ion collisions is presently under active experimental and theoretical investigation.
The  production mechanisms of hypernuclei may shed light on the theoretical understanding of the  dynamical evolution of heavy-ion reactions which cannot be addressed by other probes. In particular, 
the formation of heavy projectile/target like hypernuclei elucidates the physics at the transition region
between spectator and participant matter. 
Since hyperons are produced in the overlap region, multiplicity as well as rapidity distributions of hypernuclei formed in the target/projectile region depend crucially on the interactions of the hyperons with the hadronic matter, e.g. cross sections and potentials.
On the other hand, midrapidity hypernuclei as well as their non-strange counterparts test the phase space distribution of baryons in the expanding participant matter, especially whether the phase space distributions of strange and non-strange baryons are similar  and whether they are in thermal equilibrium.

The description of cluster and hypernuclei formation is a challenging theoretical task
which requires \\
I) the  microscopic dynamical description of the time evolution of heavy-ion collisions;\\
II) the modelling of the mechanisms for the cluster formation.\\

We study this production of hypernuclei and clusters employing
the Parton-Hadron-Quantum-Molecular-Dynamics (PHQMD) approach \cite{Aichelin:2019tnk,Glassel:2021rod}, a microscopic n-body transport model based on the QMD propagation of the baryonic degrees of freedom with density dependent 2-body potential interactions.
 In PHQMD the cluster formation occurs dynamically, caused by the interactions. The clusters are recognized by the Minimum Spanning Tree (MST) algorithm. 
We present the PHQMD results for cluster and hypernuclei formation  in comparison with the available experimental data 
at AGS, SPS, RHIC-BES and RHIC fixed target energies. 
We also provide predictions on cluster production for the upcoming FAIR experiment.
PHQMD allows to study the time evolution of formed clusters and the origin of their production, which helps to understand how such weakly bound objects are formed and survive in the rather dense and hot environment created in heavy-ion collisions. It offers therefore an explanation of the 'ice in the fire' puzzle.

To investigate whether this explanation of the 'ice in the fire' puzzle applies only to the MST results we study also the deuterons production by coalescence \cite{Kireyeu:2022qmv}. We embed MST and coalescence  in the PHQMD and UrQMD transport approaches in order to obtain  model independent results \cite{Kireyeu:2022qmv}.  We find that both clustering procedures give very similar results for the deuteron observables in the UrQMD as well as in the PHQMD environment. This confirms that our solution for the 'ice in the fire' puzzle is common to MST and coalescence  and independent of the transport approach.

\section{Cluster production in the PHQMD and UrQMD approaches}
\label{secPHQMD}

In this section we briefly recall the basic ideas of the PHQMD and UrQMD approaches
and cluster identification procedures - coalescence and MST.
For further details of the comparison of the PHSD and UrQMD dynamics we refer the readers to the recent review \cite{Bleicher:2022kcu} and on comparison
of the coalescence and MST - to the Ref. \cite{Kireyeu:2022qmv}.

The PHQMD approach \cite{Aichelin:2019tnk,Glassel:2021rod} and the Ultra-relativistic-Quantum-Molecular Dynamics (UrQMD) \cite{Bass:1998ca,Bleicher:1999xi}
describe the dynamics of heavy-ion collision on a microscopic basis using Quantum Molecular Dynamics (QMD). In QMD approaches \cite{Aichelin:1991xy,Hartnack:2015vzc} the nucleons are presented as Wigner densities of the wave functions where the nucleons interact by mutual density dependent two-body interactions between all nucleons. This allows to account for n-body nucleon correlations  (smeared out in mean-field approaches) which are important for the cluster production. 

The collision integral and description of QGP and mesonic dynamics in the PHQMD is based on the  off-shell microscopic Parton-Hadron-String Dynamics (PHSD) approach  \cite{Cassing:2008sv,Cassing:2009vt,Bratkovskaya:2011wp,Linnyk:2015rco}), based on the  Kadanoff-Baym equations \cite{KadanoffBaym} in first-order gradient expansion \cite{Cassing:2008nn}. 
The description of the QGP phase in PHSD/PHQMD is described in terms of strongly interacting quasiparticles based on the DQPM (Dynamical Quasi-Particle Model) 
 \cite{Cassing:2008nn}.
 In the PHQMD the clusters are formed dynamically, by nucleon-nucleon interactions.
 As introduced in Ref.~\cite{Aichelin:2019tnk}, 
 the cluster identification/recognition  in the PHQMD \cite{Aichelin:2019tnk}
 is done either by  the Minimum Spanning Tree (MST) procedure \cite{Aichelin:1991xy} (also used in this study) or by the Simulated Annealing Clusterization Algorithm (SACA) (see \cite{Puri:1996qv,Puri:1998te}), as it was previously done in QMD \cite{Puri:1996qv,Gossiaux:1997hp,LeFevre:2019wuj} calculations.
Clusters are identified with a MST in coordinate space, i.e when the distance between the proton-neutron pair in their rest frame is smaller than $r_0=4$ fm. We note that additional cuts in momentum space do not change much the cluster distributions found by the MST \cite{Kireyeu:2021igi}. 

UrQMD \cite{Bass:1998ca,Bleicher:1999xi}includes hadron-string  dynamics.
As degrees of freedom it includes the full set of established hadrons and their resonances, summing up to more than 50 baryon species and 40 meson species.
The UrQMD is a hybrid approach where the QGP phase can be included (optionally) into the UrQMD dynamics by switching to a hydrodynamical hybrid description \cite{Steinheimer:2007iy,Petersen:2008dd} for the most dense stages of the reaction.
We note that for the results presented here we use UrQMD in hadron-string mode without QGP, i.e. without a transition to the hydro mode. 
The light cluster formation in the UrQMD is realized by the coalescence procedure.
It has been applied to study deuteron production in the UrQMD approach for various energies, see e.g. \cite{Sombun:2018yqh,Hillmann:2019wlt,Gaebel:2020wid,Kittiratpattana:2020daw,Hillmann:2021zgj}.

We note that the coalescence procedure assumes that deuterons are created 
at freeze-out, i.e. when the last collision of the nucleons of the pair takes place and if at this time the relative distance of the two nucleons in coordinate and momentum space is lower than the coalescence parameters. 
The MST procedure can be applied any time for finding the nucleon correlations in coordinate space. This allows to follow the history of the cluster production and study their stability. As has been demonstrated in Ref. \cite{Kireyeu:2022qmv} the MST procedure asymptotically provides results which are similar to a coalescence procedure.

\section{Results on cluster production}

In this Section we present  selected results on the light cluster and hypernuclei production at relativistic energies and discuss the space-time distribution of deuteron production.
Since the clusters in the semi-classical QMD approach are not fully stable and nucleons can "spontaneously" evaporate, we show the results at different time in order to demonstrate this uncertainties (cf. Ref. \cite{Glassel:2021rod} for more details). 

\subsection{Excitation function}
\label{secEF}

\begin{figure*}[h!]
\centering
\includegraphics[width=0.32\textwidth]{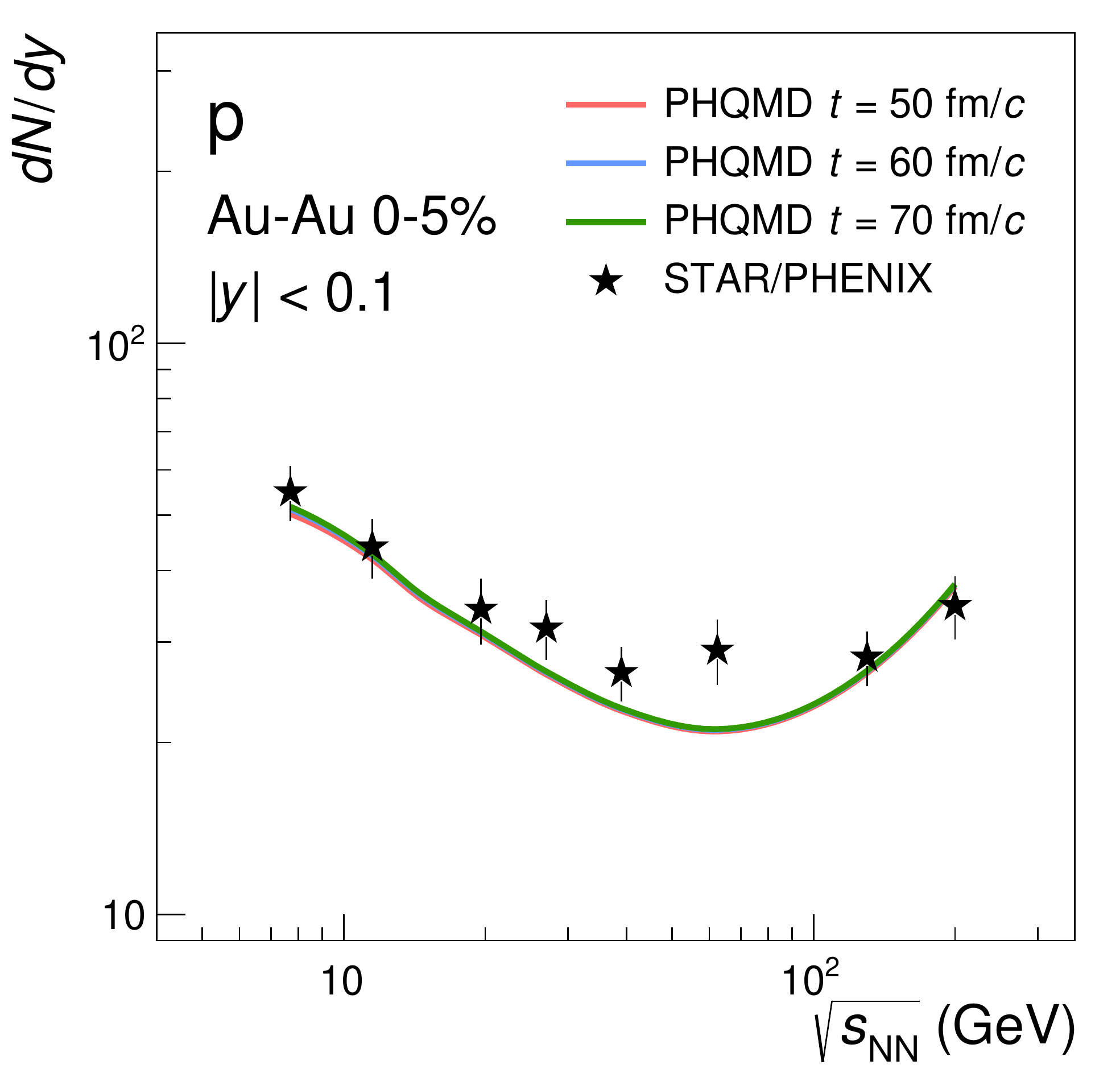}
\includegraphics[width=0.32\textwidth]{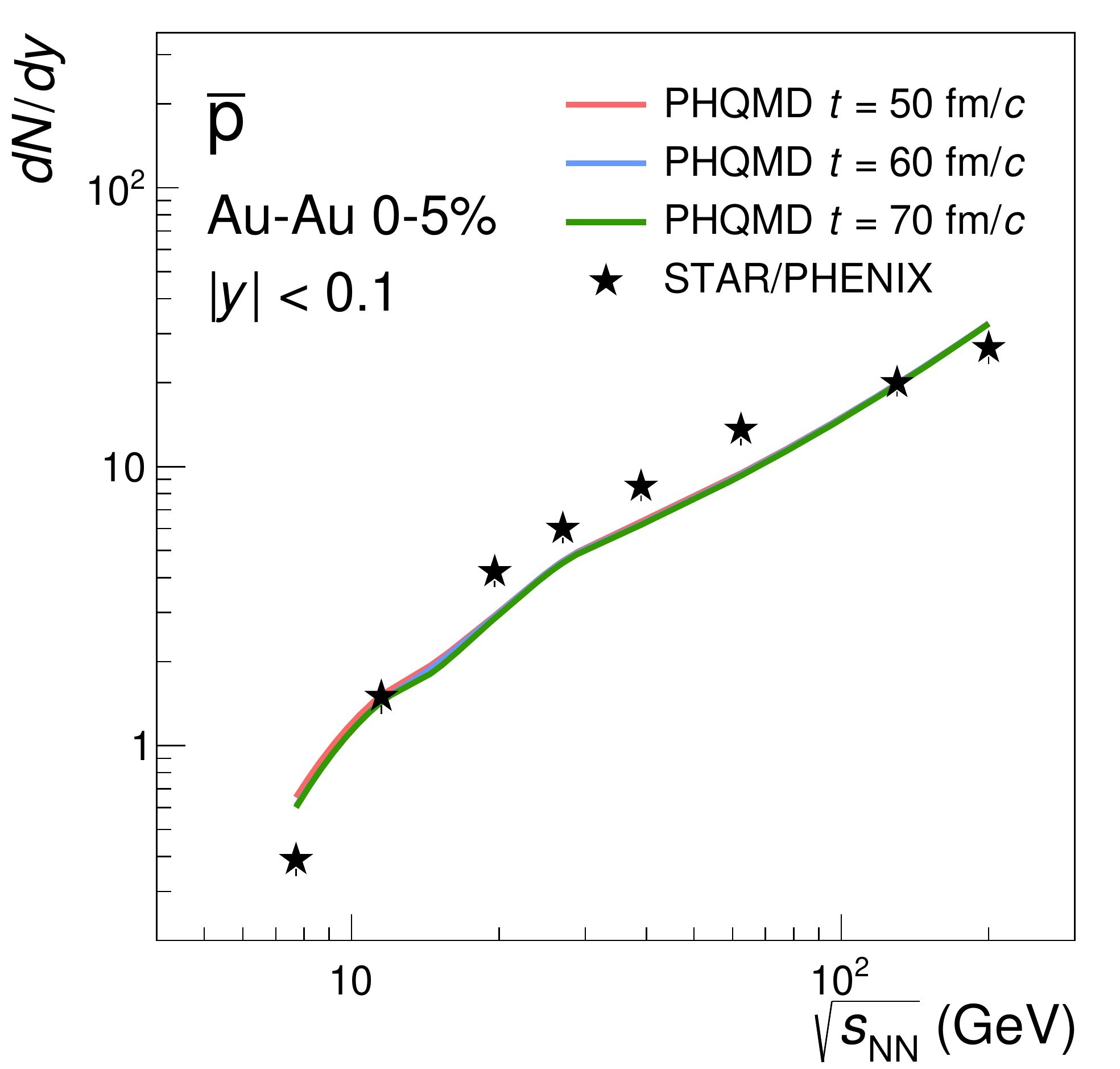}
\includegraphics[width=0.32\textwidth]{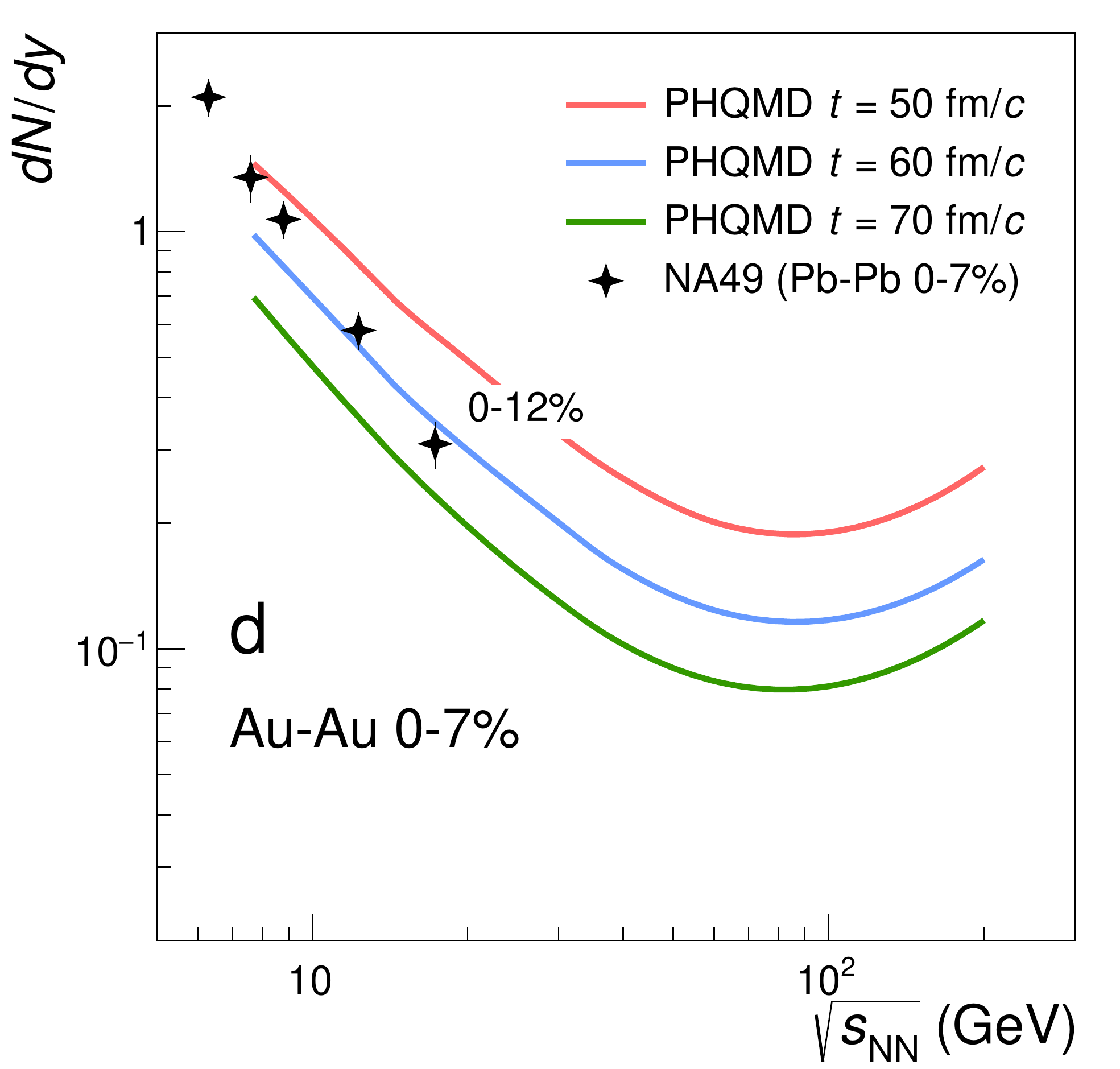}
\caption{\label{fig:WS} The midrapidity excitation function of $dN/dy$ of protons (left), antiprotons (middle) and deuterons (right)  as a function of $\sqrt{s_{NN}}$ for central Au+Au collisions (5\% most central (protons and antiprotons) and 7\% most central (deuterons) in comparison with the experimental data from the NA49 Collaboration \cite{Anticic:2016ckv} (crosses) where the midrapidity intervals are taken as $-0.4<y<0.0$ for 
$\sqrt{s_{NN}}= 6.3$~GeV, 7.6~GeV and 8.8~GeV; $-0.6<y<-0.2$ for 12.3~GeV and  $-0.6<y<-0.4$ for 17.3~GeV, 
as well as from the STAR and PHENIX Collaborations \cite{STAR:2017sal,PHENIX:2003iij} (stars) for $|y|<0.1$. The rapidity intervals for the PHQMD results correspond to the experimental ones.
The PHQMD results are shown as red lines for $t=50$ fm$/c$,  blue lines for $t=60$ fm$/c$ and green lines  for $t=70$ fm$/c$. 
The PHQMD results for protons and antiprotons  are scaled to account for the protons from weak decay  feed-down that are included in the experimental data.
The figure is taken from Ref. \cite{Glassel:2021rod}. }
\end{figure*} 

In Fig. \ref{fig:WS} we show the midrapidity excitation function of $dN/dy$ of protons (left), antiprotons (middle) and deuterons (right)  as a function of $\sqrt{s_{NN}}$ for central Au+Au collisions - 5\% most central (protons and antiprotons) and 7\% most central (deuterons) - in comparison with the experimental data from the NA49 Collaboration \cite{Anticic:2016ckv}. The PHQMD results are shown as red lines for $t=50$ fm$/c$,  blue lines for $t=60$ fm$/c$ and green lines  for $t=70$ fm$/c$. 

\begin{figure*}
\centering
\includegraphics[width=0.32\textwidth]{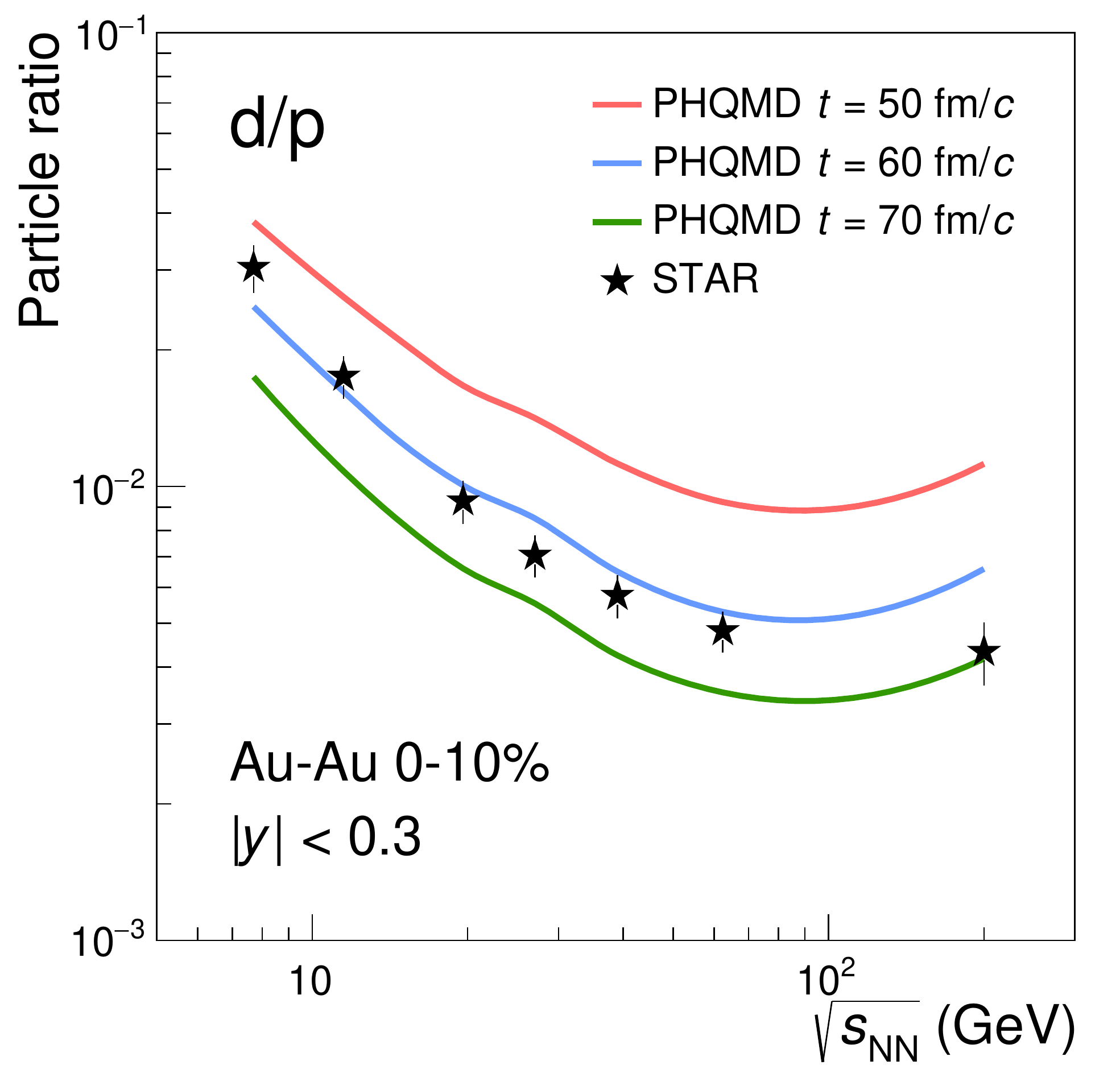}
\includegraphics[width=0.32\textwidth]{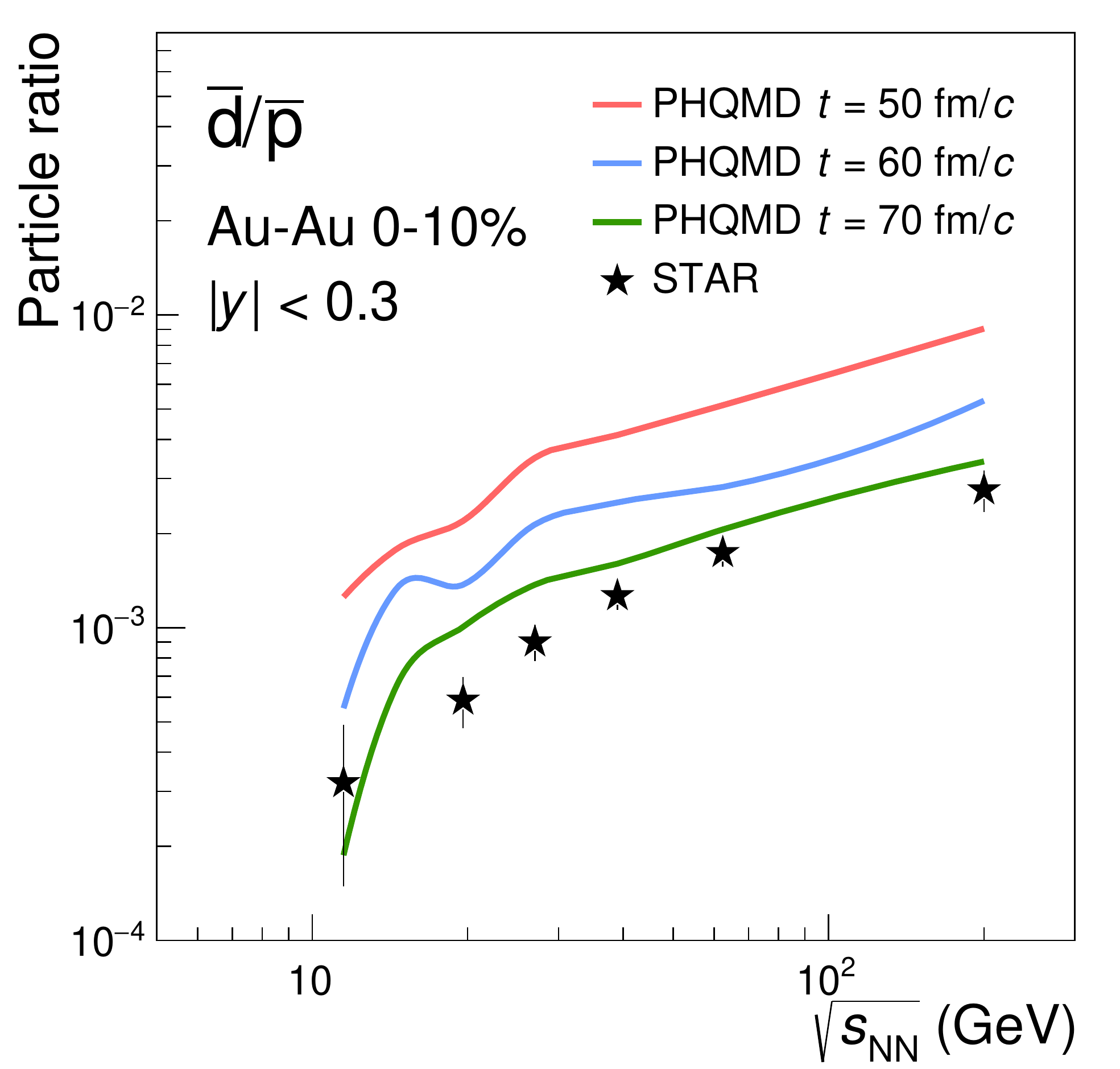}
\caption{\label{fig:rat} The excitation function of the deuteron to proton (left) and antideuteron to antiproton ratios (right) for central Au+Au collisions as a function of $\sqrt{s_{NN}}$. 
The experimental data from the STAR Collaboration \cite{Adam:2019wnb} in the rapidity interval $|y|<0.3$ are indicated as stars.
The PHQMD results are shown as red lines for $t=50$~fm$/c$, blue lines for $t=60$~fm$/c$ and green lines for $t=70$~fm$/c$. 
The figure is taken from Ref. \cite{Glassel:2021rod}. }
\end{figure*}

One can see that the proton and antiproton  as well as the deuteron $dN/dy$ are well reproduced at all energies - from the lowest SPS energies up to the highest RHIC energies. As a consequences, the experimental $d/p$ ratio versus $\sqrt{s_{NN}}$ is also reproduced rather well as follows from Fig.~\ref{fig:rat}, which shows the excitation function of the deuteron to proton $d/p$ (left) and antideuteron to antiproton $\bar d/\bar p$ (right) ratios from central Au+Au collisions at midrapidity as a function of $\sqrt{s_{NN}}$. The PHQMD results are shown as red lines for $t=50$~fm$/c$,  blue lines for $t=60$~fm$/c$ and green lines for $t=70$~fm$/c$ (similar to Fig. \ref{fig:WS}).
The form of the $\sqrt{s_{NN}}$-dependence of  $\bar d/ \bar p$  is well reproduced, however, we overestimate this ratio by about a factor of two, if the yield of antiparticles is taken at the same time as the yield of the particles. As follows from  middle part of Fig.~\ref{fig:WS} the $\bar p$ distribution does not depend on time, this enhancement is consequently due to a overpredicted $\bar d$ yield.

\subsection{Transverse spectra of light clusters }
\label{secLQ}

\begin{figure*}[h!]
\centering
\includegraphics[width=16.5cm,clip]{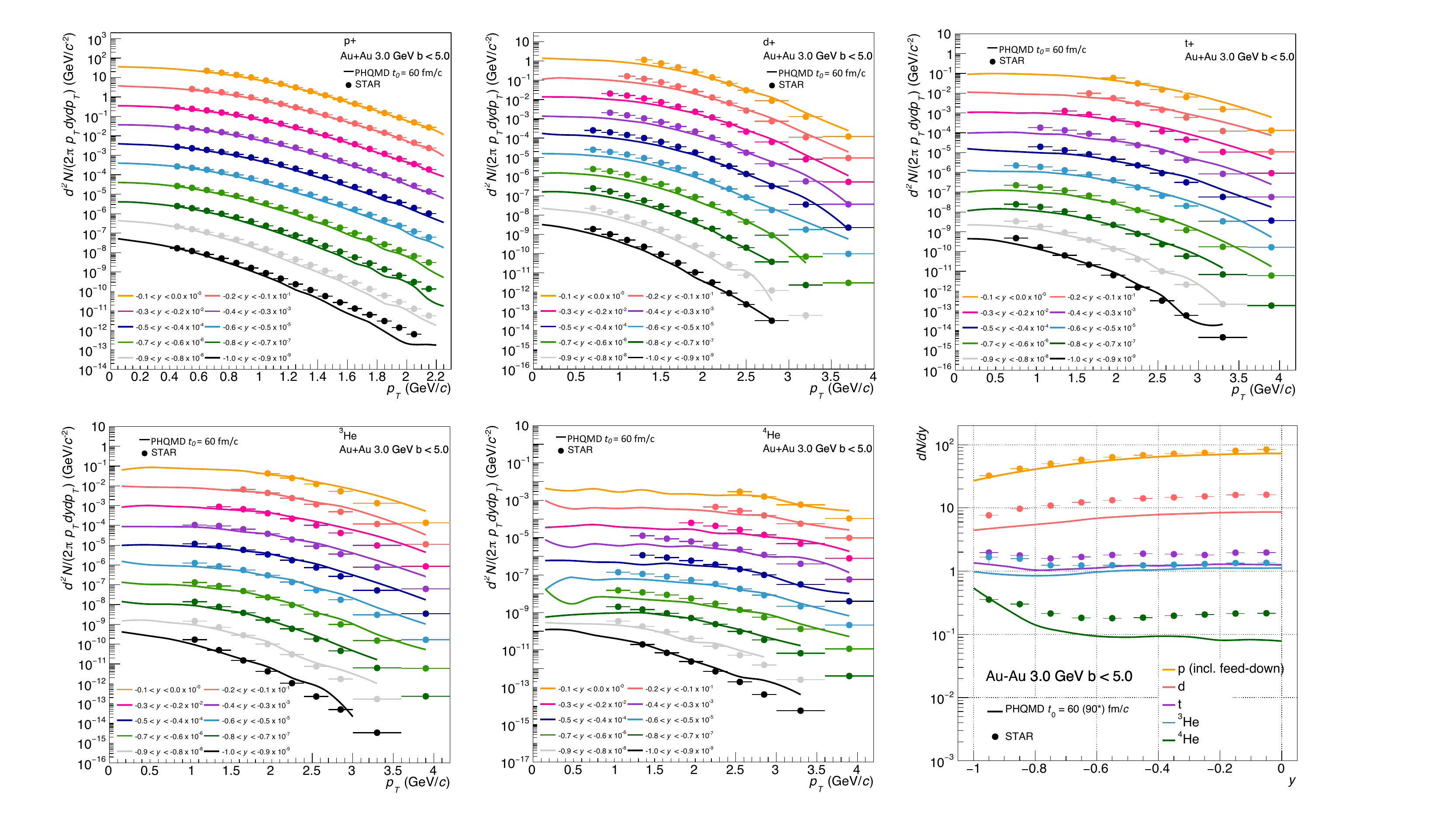}
\caption{The PHQMD results for the $p_T$ distribution of $p, d, t$, $^3He$, $^4He$ in comparison with (preliminary) STAR fixed target data \cite{STAR:3GeVpt} for Au+Au central collisions at $\sqrt{s}=3$ GeV for different rapidity bins. 
The right lower plot shows the rapidity distribution (for $y>0$) of light clusters.}
\label{fig-STAR3pt}       
\end{figure*}

In Figure \ref{fig-STAR3pt} we show the comparison of the PHQMD results with (preliminary)  fixed target STAR data \cite{STAR:3GeVpt} for the $p_T$ distribution of $p, d, t$, $^3He$, $^4He$ from Au+Au central collisions at $\sqrt{s}=3$ GeV for different rapidity bins.
The PHQMD results are shown for $t=60$~fm$/c$.
The right lower plot shows the rapidity distribution (for $y>0$) of light clusters.
One can see that the PHQMD reproduces well the $p_T$ spectra while the corresponding $y$ distribution is underestimated for the light clusters. We note that the STAR data are obtained by integration of the $p_T$ spectra extrapolated to low $p_T$ region. This 
extrapolation is different from the actual PHQMD calculations at lower $p_T$.

\subsection{Hypernuclei production}
\label{secHN}

The study of hypernuclei production can provide information 
on the local phase space density of non-strange baryons, 
as well as the phase space density of the hyperons themselves
since the hyperons are produced  at the hadronization of the QGP
and in elementary hadronic reactions during the heavy-ion collisions and then redistributed over the expanding fireball by collisions or potential interactions with nucleons.
Here we show the PHQMD results in comparison to the recent experimental data of STAR collaboration at $\sqrt{s_{NN}}= 3 $~GeV. 
\begin{figure*}[ht!]
\centering
    \includegraphics[width=0.3\textwidth]{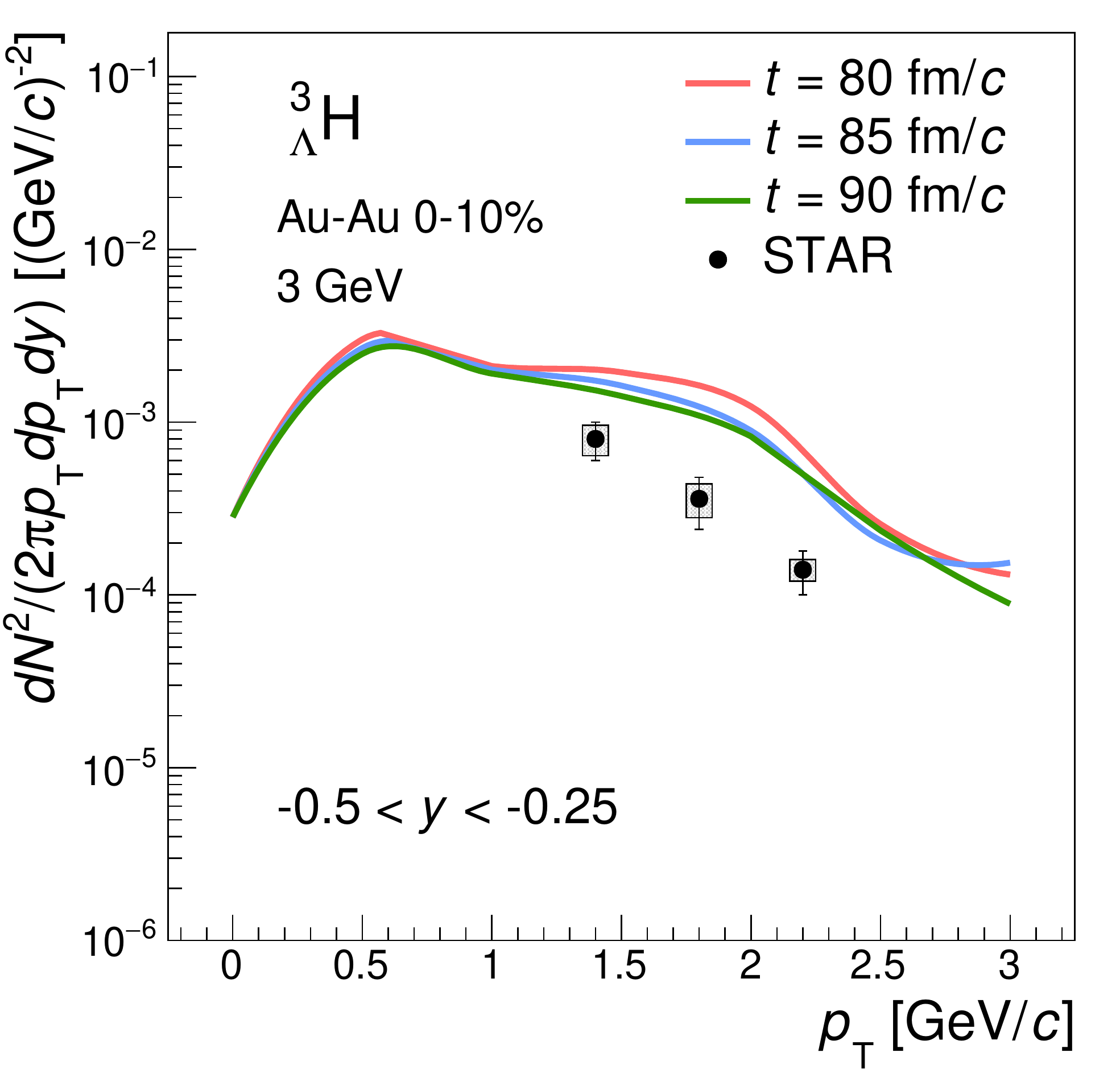}
    \includegraphics[width=0.3\textwidth]{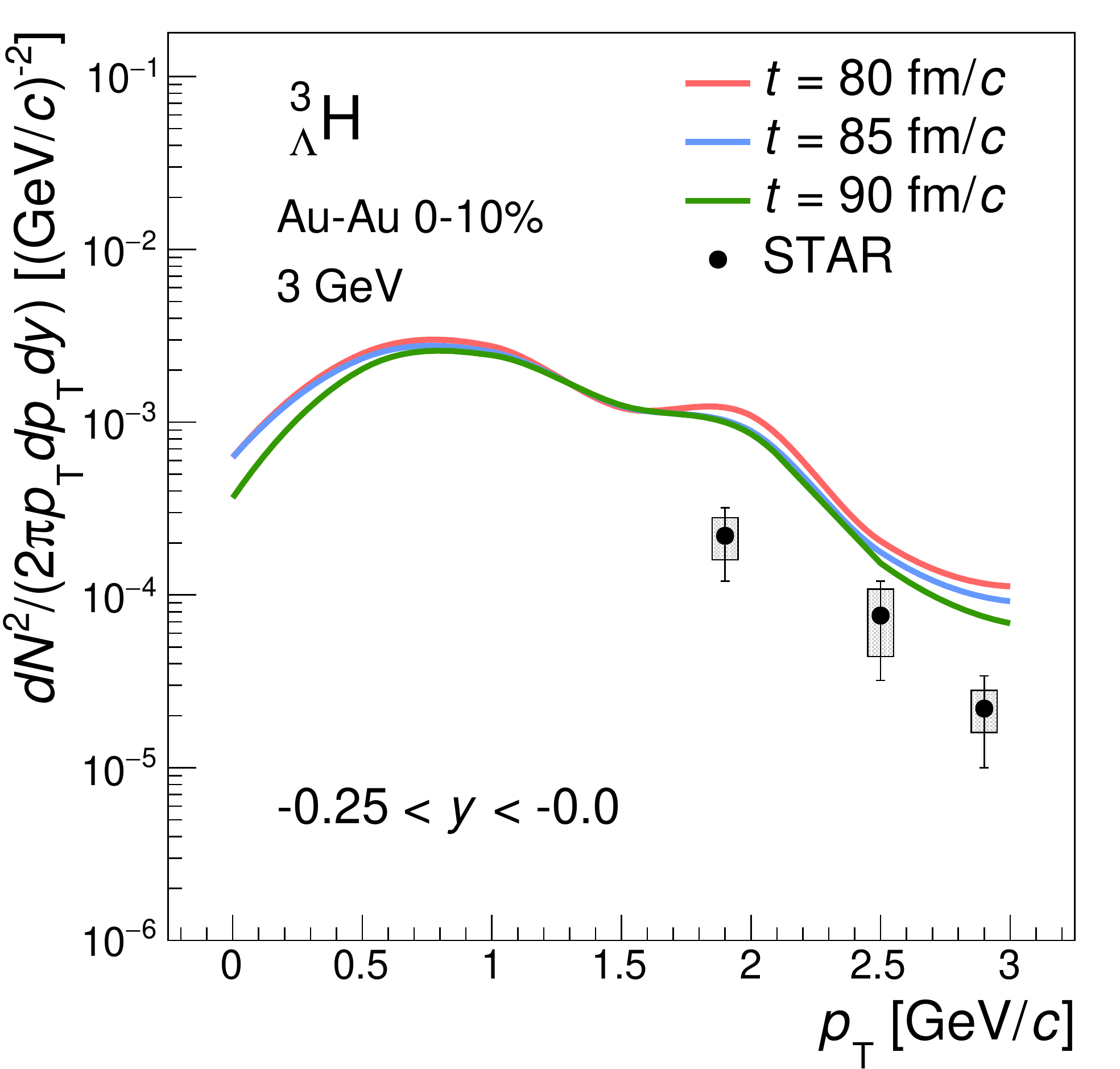}
\caption{\label{fig:H3L_pT_3} 
Transverse momentum distribution of $^{3}_{\Lambda}$H for different rapidity intervals as indicated in the legends in central Au+Au collisions at $\sqrt{s_{NN}}= 3 $~GeV. 
The filled circles indicate the experimental data from the STAR Collaboration \cite{STAR:2021orx}. The PHQMD results are taken at the times $t = 80$~fm/$c$ (red lines), 85~fm/$c$ (blue lines) and 90~fm/$c$ (green lines).
The figure is taken from Ref. \cite{Glassel:2021rod}.}
\end{figure*}
\begin{figure*}[h!]
\centering
    \includegraphics[width=0.3\textwidth]{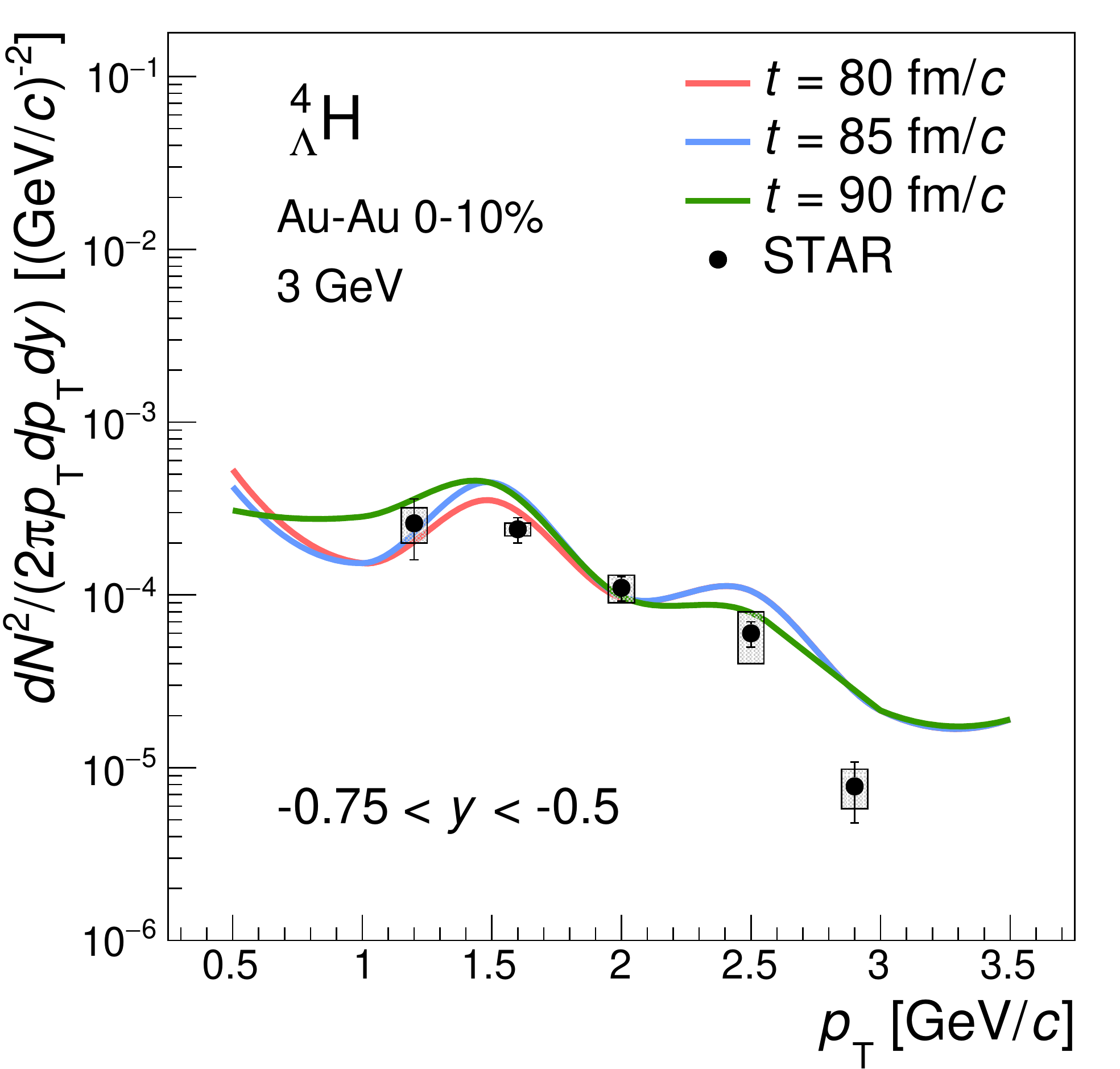}
    \includegraphics[width=0.3\textwidth]{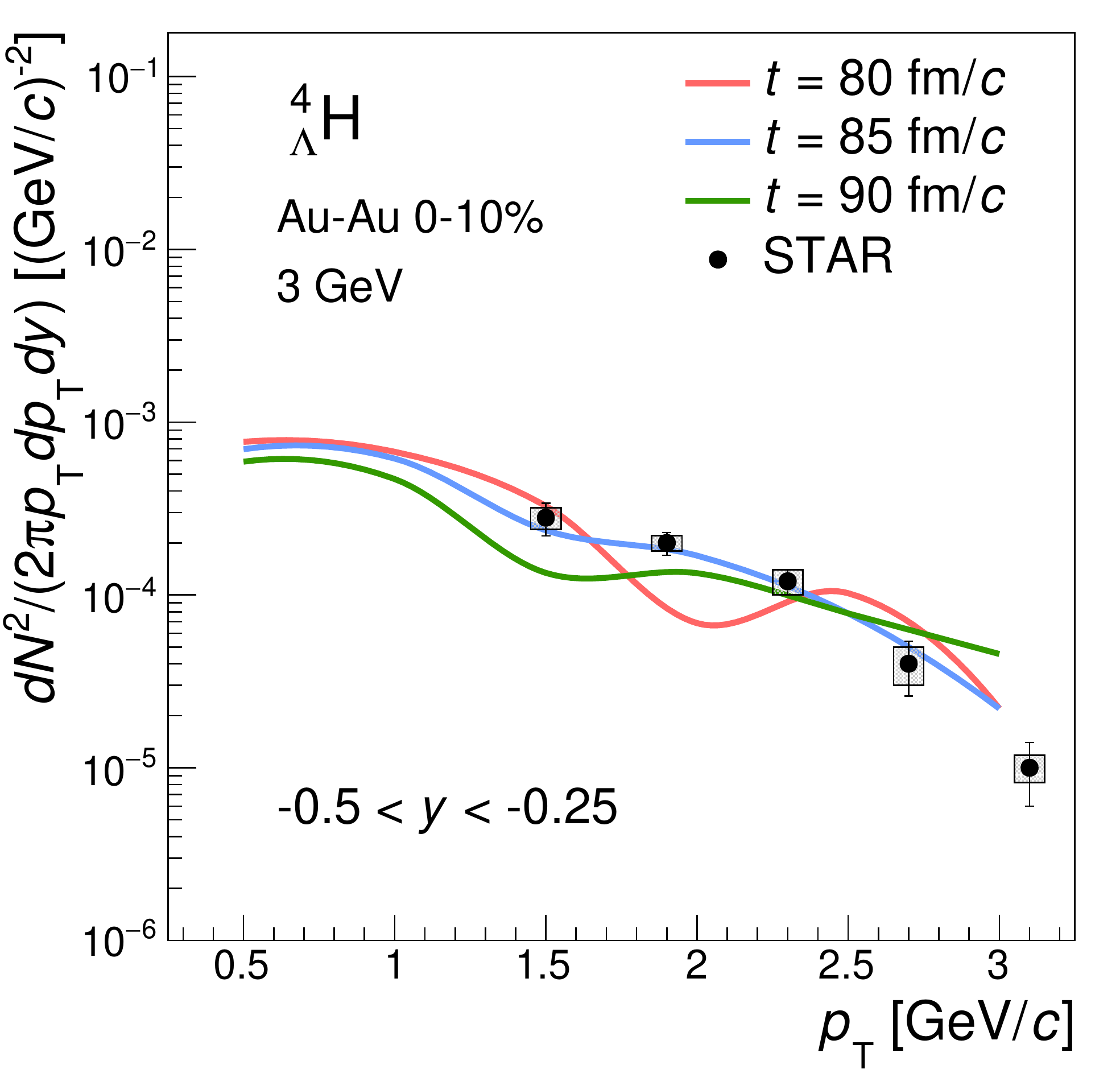}
   \includegraphics[width=0.3\textwidth]{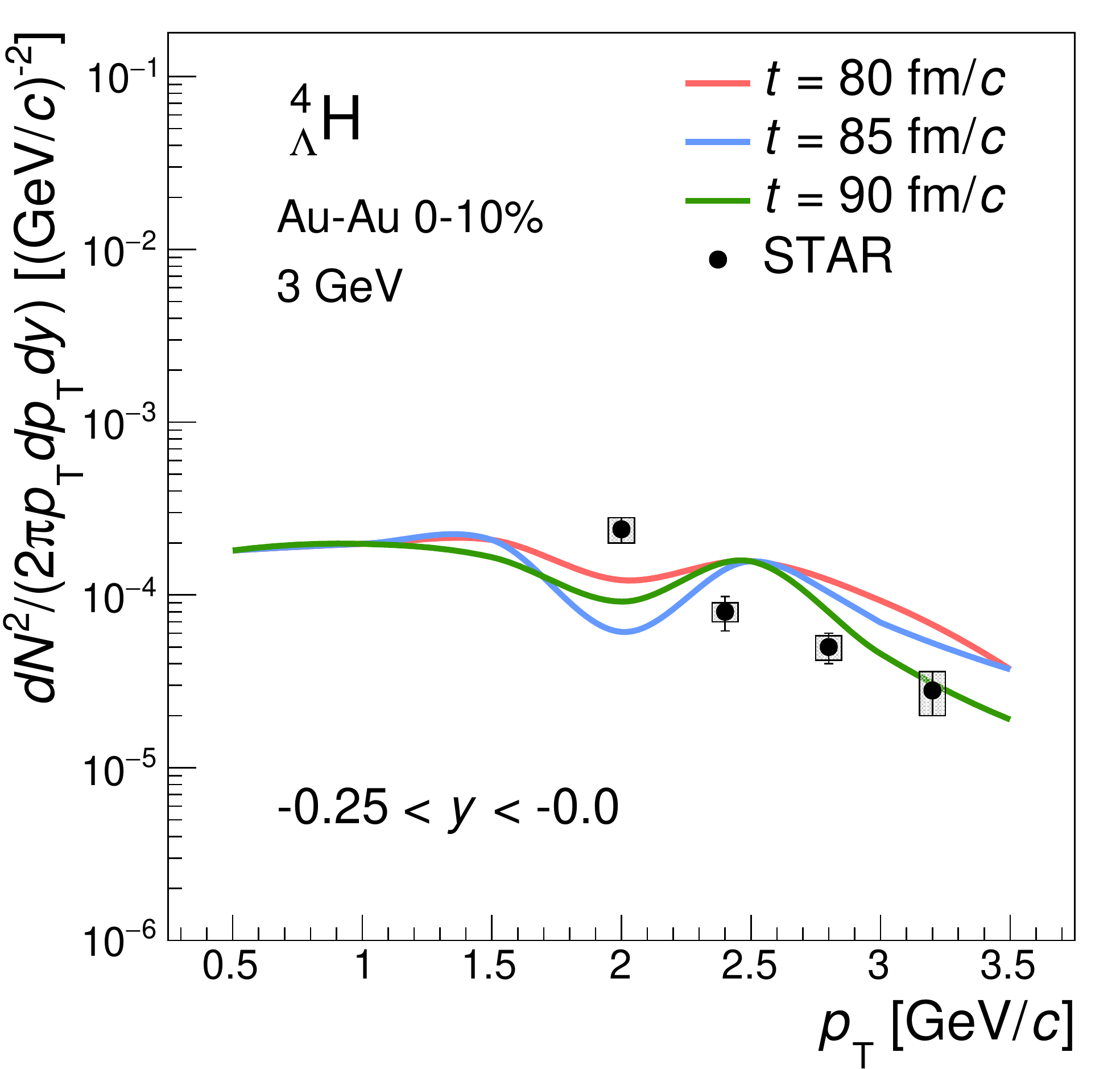}
\caption{\label{fig:H4L_pT_3} 
Transverse momentum distribution of $^{4}_{\Lambda}$H for different rapidity intervals as indicated in the legends in central Au+Au collisions at $\sqrt{s_{NN}}= 3 $~GeV. 
The filled circles indicate the preliminary experimental data from the STAR Collaboration \cite{STAR:2021orx}. The PHQMD results are taken at the times $t = 80$~fm/$c$ (red lines), 85~fm/$c$ (blue lines) and 90~fm/$c$ (green lines).
The figure is taken from Ref. \cite{Glassel:2021rod}.}
\end{figure*}

Figures~\ref{fig:H3L_pT_3} and \ref{fig:H4L_pT_3} display the comparison of the PHQMD transverse momentum distribution with the preliminary experimental data from the STAR Collaboration. In Fig.~\ref{fig:H3L_pT_3} we show the results for ${}^3_\Lambda$H, while in Fig.~\ref{fig:H4L_pT_3} those for ${}^4_\Lambda$H for different rapidity bins are presented. The filled circles indicate the experimental data from the STAR Collaboration \cite{STAR:2021orx}. The PHQMD results are taken at the times $t = 80$~fm/$c$ (red lines), 85~fm/$c$ (blue lines) and 90~fm/$c$ (green lines), but at this low energy the time dependence of the cluster yield is weak. 

The PHQMD reproduces well the shape of the experimental $p_T$ spectra. We note that the hypernuclei production is sensitive to the $\Lambda N$ potential. 
The PHQMD overestimates the differential yield of ${}^3_\Lambda H$, but reproduces the yield of ${}^4_\Lambda H$. 
This demonstrates that PHQMD is a good starting tool for more 
extended studies of the hypernucleus production in heavy-ion collisions.

\subsection{Space and time distribution of deuteron production}
\label{secST}

\begin{figure*}[th!]
    \centering
    \resizebox{\textwidth}{!}{
        \includegraphics{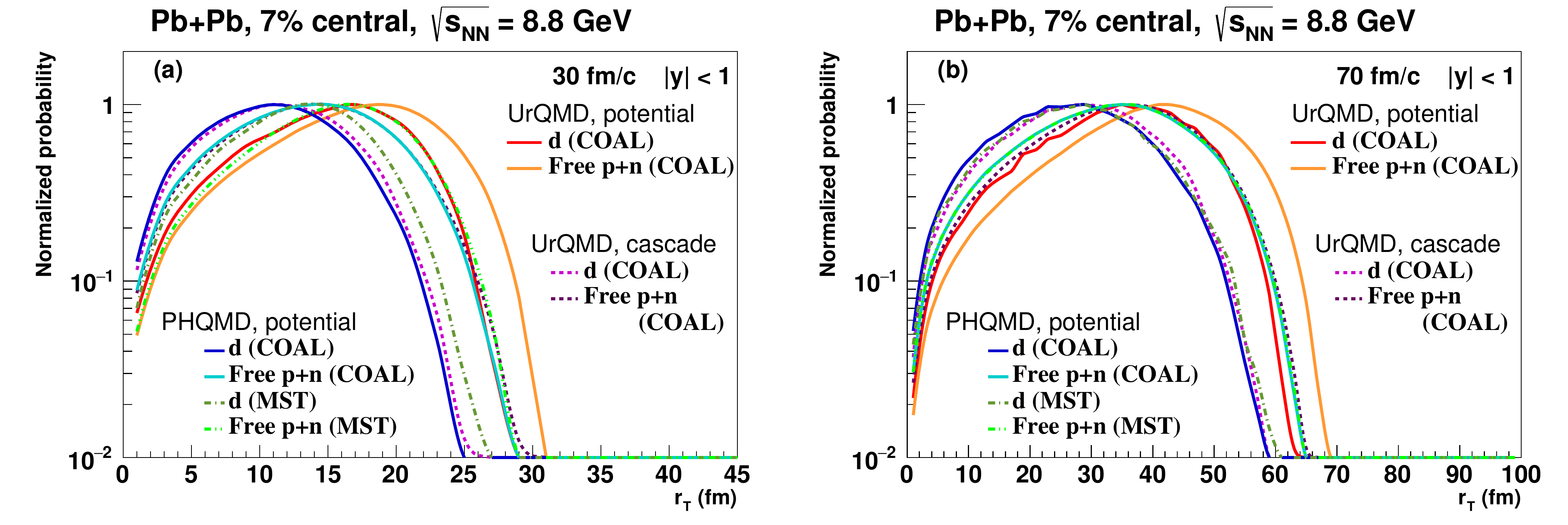}
    }
    \caption{Transverse distance of unbound nucleons ($p+n$) and deuterons at $30$ fm/c (a) and at $70$ fm/c (b).  
    The solid red line shows the deuterons from UrQMD (potential mode), 
    the solid orange line the free $p+n$ from UrQMD (potential mode), 
    the dashed magenta line represents deuterons from UrQMD (cascade mode), 
    the dashed purple line the free $p+n$ from UrQMD (cascade mode), 
    the blue solid line shows the deuterons from PHQMD (potential mode), 
    the cyan solid line the free $p+n$ from PHQMD (potential mode).
    Additionally, the dark green dot dashed line shows PHQMD deuterons found by the MST algorithm and the light green dot dot dashed line represents free $p+n$ from PHQMD with MST algorithm.
The figure is taken from Ref. \cite{Kireyeu:2022qmv}.  }
    \label{tr_dist_times}
\end{figure*}

In Ref. \cite{Kireyeu:2022qmv} we have compared two mechanisms of deuteron identification by the coalescence and MST procedures which have been implemented in two independent microscopic transport approaches - PHQMD and UrQMD.  We found that both clustering procedures give very similar results for the deuteron observables in the UrQMD and PHQMD environment. Furthermore, the PHQMD and UrQMD calculations agree well with the NA49 experimental data on deuteron production in Pb+Pb collisions at $\sqrt{s_{NN}} = 8.8$ GeV (selected for the comparison of the MST and coalescence methods and transport models in this study). 

Since the microscopic transport approaches allow to study in details the production process, the detailed investigation, performed in Ref. \cite{Kireyeu:2022qmv}, showed that the coordinate space distribution of the produced deuterons differs from that of the free nucleons and other hadrons. Thus, deuterons are not destroyed by additional rescattering. 

This finding can be demonstrated by showing  the transverse spatial distribution of nucleons and deuterons. In the left hand side of Fig.  \ref{tr_dist_times} we depict the normalized distribution at a fixed center-of-mass time of $30$ fm/c and on the right hand side at a fixed center-of-mass time of $70$ fm/c. 
 Again, the deuterons are identified by the coalescence procedure.  For PHQMD we show additionally the results for the MST procedure, which gives nearly identical results as coalescence. 
 
As follows from Fig.  \ref{tr_dist_times} the nucleon and deuteron transverse positions follow the expansion flow with the maxima of the distributions shifting by approximately $20$ fm/c in the interval between $30$ fm/c to $70$ fm/c. 
Remarkable that the deuterons remain at smaller radial distances than the nucleons; also the average $r_T$ of deuterons is in all transport approaches smaller than that of the free nucleons. 
 Consequently, the deuterons are produced at a smaller distance to the reaction center than the free protons or probably at a later time. This is a hint for a production at the end of the hadronization of the QGP or from nucleons which are closer to the center of the reaction than the average.

\section{Summary}
\label{secSum}
In this contribution we recall the selected results on midrapidity light cluster and hypernuclei  production with the PHQMD transport approach where the clusters are identified with a minimum spanning tree (MST) in coordinate space.
In QMD, being a semi-classical approach, clusters are not fully stable, thus the time at which the clusters are identified, has an influence on the multiplicity, but not on dynamical variables and is almost the same for the energy range investigated here.
The  excitation function of the light cluster yield as well as the  $p_T$ spectra
of clusters are well reproduced in a wide energy range of center of mass energies, from a few GeV up to the top RHIC energies.
    
The recent experimental data by STAR Collaboration on hypernuclei production at an
invariant energy of 3 GeV are reasonable well described.
Moreover, the cluster identification procedure by the MST has been compared with the coalescence mechanism implemented in two independent transport approaches - the UrQMD and PHQMD - for studying the deuteron production. 
The coalescence as well as the MST procedure show that the deuterons remain in transverse direction closer to the center of the heavy-ion collision than free nucleons. It offers therefore an explanation of the 'ice in the fire' puzzle:
the deuterons follow behind the front of the expanding baryonic fireball, they are spatially separated which might explain why they are not destroyed by collisions with the hadrons of the expanding fireball.

\vspace*{5mm}  
The authors acknowledge inspiring discussions with  H. Liu, C.M. Ko, I. Vasiliev.
This study is part of a project that has received funding from the European Union’s Horizon 2020 research and innovation program under grant agreement STRONG – 2020 - No 824093. Furthermore, we acknowledge support by the Deutsche Forschungsgemeinschaft 
(DFG, German Research Foundation): grants BR~4000/7-1 and BL 982/3-1
and by the GSI-IN2P3 agreement under contract number 13-70. 
JS acknowledges support from the BMBF through the ErUM-data-pilot project as well as the Samson AG.
Further support was provided by the PPP Program of the DAAD.
Computational resources were provided by the Center for Scientific Computing (CSC) of the Goethe University, by the "Green Cube" at GSI, Darmstadt.

\bibliography{biblio-SQM}

\end{document}